\documentclass[11pt, reqno]{amsart}
\usepackage{amsmath, amsthm, a4, latexsym, amssymb}

\def\@rmrk#1#2{\refstepcounter
    {#1}\@ifnextchar[{\@yrmrk{#1}{#2}}{\@xrmrk{#1}{#2}}}

\makeatletter\@addtoreset{equation}{section}\makeatother

\newfont{\bfit}{cmbxti10 scaled 2000}
\newfont{\biggi}{cmr12 scaled 2000}
\newtheorem{step}{STEP}

\newcommand{\bes}{\begin{step}}
\newcommand{\es}{\end{step}}
 
 \newcommand{\eps}{\varepsilon}

 \newcommand{\prob}{\mathbb{P}}

 \renewcommand{\P}{\mathbb{P}}
 
 \newcommand{\skrie}{{\mathcal E}}
 \newcommand{\skrif}{{\mathcal F}}
 \newcommand{\skrig}{{\mathcal G}}
 \newcommand{\skrih}{{\mathcal H}}

 \newcommand{\skril}{{\mathcal L}}
 \newcommand{\skrim}{{\mathcal M}}

 \newcommand{\skrix}{{\mathcal X}}

 \newcommand{\sfrac}[2]{\mbox{$\frac{#1}{#2}$}}

\def\1{{\mathchoice {1\mskip-4mu\mathrm l}      
{1\mskip-4mu\mathrm l}
{1\mskip-4.5mu\mathrm l} {1\mskip-5mu\mathrm l}}}

\newcommand{\eq}{\begin{equation}}
\newcommand{\en}{\end{equation}}

\renewcommand{\subsection}{\secdef \subsct\sbsect}
\newcommand{\subsct}[2][default]{\refstepcounter{subsection}
\vspace{0.15cm}
{\flushleft\bf \arabic{section}.\arabic{subsection}~\bf #1  }
\nopagebreak\nopagebreak}
\newcommand{\sbsect}[1]{\vspace{0.1cm}\noindent
{\bf #1}\vspace{0.1cm}}

\newtheorem{theorem}{Theorem}[section]
\newtheorem{lemma}[theorem]{Lemma}

\newtheoremstyle{thm}{1.5ex}{1.5ex}{\itshape\rmfamily}{}
{\bfseries\rmfamily}{}{2ex}{}

\newtheoremstyle{rem}{1.3ex}{1.3ex}{\rmfamily}{}
{\itshape\rmfamily}{}{1.5ex}{}
\theoremstyle{rem}

\refstepcounter{subsection}

\def\thebibliography#1{\section*{reference}
  \list%
  {\arabic{enumi}.}
    {\settowidth\labelwidth{[#1]}\leftmargin\labelwidth
    \advance\leftmargin\labelsep
    \parsep0pt\itemsep0pt
    \usecounter{enumi}}
    \def\newblock{\hskip .11em plus .33em minus .07em}
    \sloppy                   
    \sfcode`\.=1000\relax}

\begin{document}
\title[AEP for WSN]
{\Large Large  deviation, Basic information theory  for  Wireless  Sensor  Networks}

\author[Kwabena Doku-Amponsah]{}

\maketitle
\thispagestyle{empty}
\vspace{-0.5cm}

\centerline{\sc{By Kwabena Doku-Amponsah}}
\renewcommand{\thefootnote}{}
\renewcommand{\thefootnote}{1}
\renewcommand{\thefootnote}{}
\footnote{\textit{Address:} Statistics Department, University of
Ghana, Box LG 115, Legon,Ghana.\,
\textit{E-mail:\,kdoku@ug.edu.gh}.}
\renewcommand{\thefootnote}{1}
\centerline{\textit{University of Ghana}}

\begin{quote}{\small }{\bf Abstract.}
In this  article,  we prove  Shannon-MacMillan-Breiman Theorem  for  Wireless  Sensor  Networks modelled  as coloured geometric random graphs.  For large $n,$ we  show  that a Wireless Sensor  Network  consisting of $n$ sensors  in  $[0,1]^d$ connected by an average number of links of order $n\log n $ can be coded by about $[n(\log n )^2\pi^{d/2}/(d/2)!]\,\skrih $\, bits, where $\skrih$ is an explicitly defined entropy.
In  the  process, we  derive  a joint  large  deviation  principle (LDP) for the \emph{empirical sensor measure} and \emph{the
empirical link measure} of  coloured random geometric graph models.
\end{quote}\vspace{0.5cm}

\section{Introduction}
Wireless  Sensor  Networks  (WSN)  are  now  popular  among  theoretical  computer  scientists  because  of  its  uses  as  a  tool  for  monitoring  and  controlling  the  physical  environment. Many  researches  on  finding  a  good network model  for  WSN  have  suggested  models  that  have  their  origin  in  classical  areas  of  theoretical  computer  and  applied  mathematics: regardless  of  the  radio  technology  use from  the  topology  point  of  view, at  any  instant  in  time  a  WSN  can  be  represented  as  a  graph  with  a  set  of  vertices   consisting of  nodes   of  the  network  and  a  set  of  edges  consisting  of  the  links  between  nodes.   However, recent  studies,  see Kenniche \cite{KH10}  and  the  references therein,  have  shown  that  a  random  strategy is  the  only  way  to deploy  the  large  number  of  sensors  in  inaccessible  areas  and  the  random geometric  graph  or  geometric  random  graph is  the most appropriate model. Finding a  good  coding  schemes  and  approximate pattern  matching  algorithms will  be  vital  for coding  the  WSN,  and  the Shannon-MacMillan-Breiman  Theorem (SMBT) or Asymptotic Equipartition  Property  will  be  of  great  help  in  this  regard. See,  example Dembo and~Kontoyiannis \cite{DK02}.\\

In Information theory, the  Shannon-MacMillan-Breiman  Theorem (SMBT)  or Asymptotic Equipartition  Property  is  the  analog of  the strong  law of  large  numbers.   It  is  a  direct  consequences  of  the  weak  law  of  large  for appropriately defined  statistics  of  a  stochastic  data  source. It  allows us  to partition output sequence   of  a  stochastic data  source  into  two  sets,  the  typical  set,  where  the  sample  entropy is  close  to  the  true  entropy,  and  the  non-typical set, which contains  the  other  sequence.  See, Cover  and~Thomas~ \cite{CT91}.

In  this  article  we  derive  the  SMBT  for  WSN  modelled  as  CGRG models,  using  some  of  the  large  deviation  techniques  developed  for  studying  information  theory,  see \cite{DA10},  for  networked  data  structures.\\

To be specific  we  derive  joint  large  principle  for  the  empirical sensor  measure and  the empirical  link  measure   of  the  CGRG  using \cite[Theorem~3.3]{DA10}  and the  methods  developed  therein.  From  this  LDP  we  prove  the  weak law  of  large  numbers  for  the empirical sensor  measure and  the empirical  link  measure. From  the  weak  law  of  Large  numbers  we  derive  the  SMBT for  CGRG as a  model for  the  WSN.\\

\subsection{The coloured  geometric  random  graph model.}  In this subsection  we shall describe  a
more general model of random  geometric graphs, the CGRG in
which the connectivity radius depends on  the type or colour or
symbol or  spin or sensor of the  nodes. The empirical sensor measure and the
empirical link measure are our main object of study  here.

Given   a probability measure $\nu$ on $\skrix$ and a function
$r_n\colon\skrix\times\skrix\rightarrow (0,1]$ we may define the
{\em randomly coloured  geometric random graph} or simply
\emph{coloured random geometric graph}~$X$ with $n$ vertices as
follows: Pick sites  $Y_1,...,Y_n$  at  random  independently
according to the uniform distribution on $[0,\,1]^d.$ Assign to each
site $Y_j $ sensor $X(Y_j )$ independently according to the {\em
sensor law} $\mu.$ Given the sensors, we join any two vertices
$Y_i,Y_j $,$(i\not=j)$ by a link independently of everything else,
if $$\|Y_i-Y_j \|\le r_n\big[X(Y_i),X(Y_j )\big].$$ In this  article
we  shall  refer to $r_n(a,b),$   for $a,b\in\skrix$ as a connection
radius,  and always consider
$$X=((X(Y_i),X(Y_j ))\,:\,i,j=1,2,3,...,n),E)$$ under the joint law of
graph and sensor. We interpret $X$ as  CGRG with vertices
$Y_1,...,Y_n$ chosen at  random uniformly   and independently from
the vertices space $[0,1]^d.$ For  the  purposes of  this  study we
restrict ourselves to  the sparse, intermediate and dense cases .i.e. the
connection radius $r_n$ satisfies the condition $n r_n^d(a,b)/\log n \to
\lambda_{[d]}(a,b)$ for all $a,b\in \skrix$, where $\lambda\colon\skrix^2\rightarrow
[0,\infty)$ is a symmetric function, which is not identically equal
to zero.  The  CGRG  have  been  suggested  by  Cannings  and  Penman \cite{CP03}  as  a  possible  extension  to  the  coloured  random  graphs  introduced  in Penman\cite{Pe98}.

The  distance  $r_n$  plays  a  role  similar  to  that  of  $p_n$  in  the  coloured  random  graph model  proposed  in Penman~\cite{Pe98} and  studied  by Doku-Amponsah\cite{DA06}.  Based  on  one's  choice  of  $r_n,$ qualitatively,  different  types  of  behaviour  can  be  seen.Note  that,  intuitively, the   the  average  degree scales  with  $nr^d.$ To  be  more  specific,  it  can  be  show  that  in te  classical  random  geometric graph the  ratio of  the  average  degree divided  by  $nr^d$  tends  to  a constant  in  probability  as  long  as  $n^2r_n^d\to\infty.$  See, \cite{MM05}. As  a  result  of  the  interpretation of  $nr_n^d$    a  measure  of  the  average  degree,  we  refer to  the  case  where $nr^d/log~n\to \lambda_{[d]}= 0$ as sparse  case,  the  case $nr^d/log~n\to \lambda_{[d]}$  as  the intermediate  case(s)  and  $nr^d/log~n\to \lambda_{[d]}=\infty$ as the  dense  case.\\

We associate with any coloured graph $X$ a probability measure, the
\emph{empirical sensor measure}~$\skril_X^1\in\skrim(\skrix)$,~by
$$\skril_X^{1}(a):=\frac{1}{n}\sum_{j=1}^{n}\delta_{X(Y_j )}(a),\quad\mbox{ for $a_1\in\skrix$, }$$
and a symmetric finite measure, the \emph{empirical link measure}
$\skril_X^{2}\in\tilde\skrim_*(\skrix^2),$ by
$$\skril_X^{2}(a,b):=\frac{1}{n\log n}\sum_{(i,j)\in E}[\delta_{(X(Y_i),X(Y_j ))}+
\delta_{((X(Y_j ),X(Y_i ))}](a,b),\quad\mbox{ for $(a, b)\in\skrix^2$.
}$$ The total mass $\|\skril_X^2\|$ of the empirical link measure is
$2|E|/n\log n$.

For any finite or countable set $\skrix$ we denote by
$\skrim(\skrix)$ the space of probability measures, and by
$\tilde\skrim(\skrix)$ the space of finite measures on $\skrix$,
both endowed with the weak topology.  

\section{Statement of main results}\label{AEP}

Through out  the  remaining part  of  this  article  we  assume  $d\ge 2$ is finite.

\subsection{Asymptotic Equipartition Property the Sparse and Intermediate for WSN}

The  underlying  question is,  how  many bits  are  needed  to  store  or  transmit the  information  contained  in  a  Wireless  Sensor Network consisting  of  $n$  \emph{sensors}   connected  by  number  of \emph{links}?  This  question  can  be  answered  by the   SMBT  for   Wireless  Sensor Networks,  see Theorem~\ref{randomgsmb.sparse}. To  the  SMBT we denote by $P$ the distribution of the an CGRG.  We define the measure
$\lambda_{[d]}\omega\otimes\omega\in\tilde\skrim(\skrix\times\skrix)$
by $$\lambda_{[d]}\omega\otimes\omega(a,b)=\lambda_{[d]}(a,b)\omega(a)\omega(b),\mbox{ for
$a,b\in\skrix$ }$$ and  write  $$\int_{\skrix^2} \lambda_{[d]}\omega\otimes\omega(da,db)=:\sum_{a,b\in\skrix} \omega(a) \lambda_{[d]}(a,b) \omega(b).$$

\begin{theorem}\label{randomgsmb.sparse}\label{CGRGldpg}
Suppose that $X$ is an CGRG with sensor law $\nu$ and
connection radius $ r_n\colon\skrix^2\rightarrow[0,1]$ satisfying $n
r_n^d(a,b)/log~n \to \lambda_{[d]}(a,b),$ for some symmetric function
$\lambda_{[d]}\colon\skrix^2\rightarrow [0,\infty)$ not identical to zero.
Then, for every $\eps>0$,
$$\lim_{n\rightarrow\infty}\prob\Big\{\Big|-\sfrac{1}{ n(\log n)^2}\log
P(X)-\sfrac{\pi^{d/2}}{2(d/2)!}\int_{\skrix^2} \lambda_{[d]}\nu\otimes\nu(da,db) \Big|\ge
\eps\Big \}=0.$$
 \end{theorem}

In other words, in order to transmit an WSN  in
the given sparse or  intermediate regime one needs with high probability, about
$\,[n (\log n)^2\pi^{d/2}/(d/2)!] \skrih \,\,\mbox{bits,}$  where  $\skrih$  is  the entropy  defined  by  $$\skrih:=\sfrac{1}{2 \log 2}\int_{\skrix^2} \lambda_{[d]}\nu\otimes\nu(da,db).$$
For  the  pair  of  measures  $(\omega,\varpi)$ we  define  the  near entropy  ${\mathfrak{H}_{\lambda_{[d]}}}$ by
$${\mathfrak{H}_{\lambda_{[d]}}}(\varpi\, \| \, \omega ):=
H\big(\varpi\,\|\,\rho(d)\lambda_{[d]}\omega\otimes\omega\big)+\rho(d)\|
\lambda_{[d]}\omega\otimes\omega \| -\|\varpi\|\, ,$$

where  $H\big(\varpi\,\|\,\tilde{\varpi}\big)$  means the relative  entropy  of  the  finite measure $\varpi$  with  respect  to $\tilde{\varpi}.$

\subsection{Large-deviation principles in the Sparse and Intermediate CGRG.}
The  following  LDP  is  a  key  ingredient in  the  proof  of  our  SMBT,  see Theorem~\ref{randomgsmb.sparse}

\begin{theorem}\label{randomge.jointL2L1L1d}\label{CGRGldp1}
Suppose that $X$ is an  CGRG with sensor law
$\nu\colon\skrix\rightarrow (0,1]$ and connection radius
$r_{n}:\skrix\times\skrix\rightarrow[0,1]$ satisfying
$nr_{n}^d(a,b)/\log n\rightarrow \lambda_{[d]}(a,b)),$  with
$\lambda_{[d]}:\skrix\times\skrix\rightarrow[0,\infty)$ symmetric.
 Then, for  $n\rightarrow\infty,$ the pair $(\skril_X^1,\skril_X^2)$
satisfies  a large deviation principle in
$\skrim(\skrix)\times\tilde{\skrim}_{*}(\skrix\times\skrix)$ with speed
\begin{itemize}
\item[(i)]  $n \,log\, n$ and  good rate function,
\begin{equation}\label{randomge.rateL2L1L1ns}
I_{1}^{[d]}(\omega,\varpi)=\sfrac{1}{2}{\mathfrak{H}_{\lambda_{[d]}}}(\varpi\,\|\,\omega),
\end{equation}

whilst   $\mathfrak
H_C^{d}(\varpi\,\|\,\omega)\ge 0$ and equality holds if and only if
$\varpi=\rho(d) \lambda_{[d]}\omega\otimes\omega$.

\item[(ii)] $n$ and  good rate function,
\begin{align}\label{randomge.rateL2L1L1nd}
I_{2}^{[d]}(\omega,\varpi)=\left\{
  \begin{array}{ll}H(\,\omega\,\|\,\nu\,) & \mbox { if  $\varpi=\rho(d)\lambda_{[d]}\omega\otimes\omega$, }\\
\infty & \mbox{otherwise.}
\end{array}\right.
\end{align}
\end{itemize}

\end{theorem}

\section{Derivation   of  Theorems~\ref{randomge.jointL2L1L1d} and~\ref{CGRGldpg}  }

For  any two  points  $U_1$  and  $U_2$  uniformly  and
independently chosen  from  the  space $[0,\,1]^d$  write
$$F(t):=\P\Big\{\|U_1-U_2\|\le t\Big\},$$

 where $F(r_n(a,b))=\rho(d)r_n^d(a,b),\, a,b\in\skrix^2$ i.e. the  volume of a
$d$-dimensional (hyper)sphere with radius $r(a,b)$ satisfying
$nr_n^d (a,b)/\log n\to \lambda_{[d]}(a,b)).$ Let $p_n(a,b)=F(r_n(a,b))=\rho(d)r_n^d(a,b)$  and  $$C(a,b))=\lambda_{[d]}(a,b).$$ Then    we have

\begin{align}
d\prob(X)&= \prod_{u\in V}\nu(X(Y_u))\prod_{(u,v)\in E}F(r_n(X(Y_u),X(Y_v)))\prod_{(u,v)\not\in E}1-F(r_n(X(Y_u),X(Y_v)))\\
&= \prod_{u\in V}\nu(X(Y_u))\prod_{(u,v)\in E}p_n(X(Y_u),X(Y_v))\prod_{(u,v)\not\in E}1-p_n(X(Y_u),X(Y_v))=d\tilde{\P}(X),
\end{align}
where  $\tilde{\P}(X)$  is  the  law  of  coloured  random graph  $X$ with the  geometric plane  $[0,1]^d$ ignored.

Hence  by  the  exponential  equivalence,  see \cite[Theorem~4.2.13]{DZ98} and \cite[Theorem~3.3]{DA10}   we  have  Theorem~\ref{CGRGldpg}  with  rate  functions  $I_1^{[d]}$ and  $I_2^{[d]}.$

\subsection{Derivation of Theorems~\ref{randomgsmb.sparse}}\label{DMT}\label{PAEPSHS}
\begin{lemma}\label{WLLN}
Suppose that $X$ is an CGRG with sensor law
 $\nu\colon\skrix\rightarrow (0,1]$ and
 connection radius $r_{n}:\skrix\times\skrix\rightarrow[0,1]$ such
 that $nr_n^d(a,b)/log ~n \to \lambda_{[d]}(a,b),$ for $\lambda_{[d]}:\skrix\times\skrix\rightarrow[0,\infty)$ nonzero. Then, for
 any $\eps>0$ we have

$$\lim_{n\to\infty} \prob\big\{ \sup_{a\in\skrix} |\skril_X^1(a) - \nu(a)|
 \ge \eps \big\}=0 $$  and

 $$\lim_{n\to\infty} \prob\big\{ \sup_{a,b\in\skrix} |\skril_X^2(a,b) - \nu(a) \lambda_{[d]}(a,b)) \nu(b) |\ge \eps \big\}=0.$$
 \end{lemma}

From Theorem~\ref{randomge.jointL2L1L1d}(ii), we prove this lemma.  To begin, we define a closed set $$F_1=\big\{
(\omega,\varpi)\in\skrim(\skrix) \times
\tilde{\skrim}_{*}(\skrix\times\skrix)\colon \sup_{a,b\in\skrix}
|\varpi(a,b) - \nu(a) \lambda(a,b)) \nu(b) | \ge \eps\}$$  and  $$F_2=\big\{
(\omega,\varpi)\in\skrim(\skrix) \times
\tilde{\skrim}_{*}(\skrix\times\skrix)\colon \sup_{a\in\skrix}
|\omega(a) - \nu(a) | \ge \eps\}.$$

 We observe that, by Theorem~\ref{randomge.jointL2L1L1d}(ii),
\begin{equation}\label{randomsmb.lastb1}
\limsup_{n\rightarrow\infty}\sfrac{1}{n}\log\prob\Big\{(\skril_X^1,\skril_X^2) \in
F \Big\}\le -\inf_{(\omega,\varpi)\in \skrif} I_2^{[d]}(\omega,\varpi),
\end{equation}
where  $\skrif=F_1\cup F_2.$

This will  be  shown by contradiction that the right handside of
\eqref{randomsmb.lastb1} is negative. For this purpose,  we suppose that
there exists sequence  $(\omega_n,\varpi_n)$ in $\skrif$ such that
$I_2^{[d]}(\omega_n,\varpi_n)\downarrow 0.$ Then, because $I_2^{[d]}$ is a good rate
function and its level sets are compact, and by  lower
semi-continuity of the mapping $(\omega,\varpi)\mapsto
I_2^{[d]}(\omega,\varpi)$, there is a limit point $(\omega,\varpi)\in \skrif$
with $I_2^{[d]}(\omega,\varpi)=0$. By Theorem~\ref{randomge.jointL2L1L1d}(i),
we have $H(\omega\,\|\,\nu)=0$ and
${\mathfrak{H}_C}(\varpi\,\|\,\omega)=0.$ This implies
$\omega(a)=\nu(a)$ and $\varpi(a,b)=\lambda_{[d]}(a,b))\omega(a)\omega(b),$ for
$a,b\in\skrix$ which contradicts $(\omega,\varpi)\in \skrif$. Hence
 as desired.\\

Recall that  $V$ is  a fixed set of $n$ vertices, say
$V=\{1,\ldots,n\},$  $\skrig_n$ is the set of all (simple) graphs
with vertex set $V=\{1,\ldots,n\}$ and edge set
$E\subset\skrie:=\big\{(u,v)\in V\times V \, : \, u<v\big\}.$  Now we compute the distribution
$P_n\colon\skrig_{n}(\skrix)\rightarrow[0,1]$ of $X,$
\begin{align*}
P(x)& 
=\prod_{u\in V}\nu(x(y_u))\prod_{(u,v)\in
E}F(r_{n}(x(y_u),x(y_v)))\prod_{(u,v)\not\in
E} \big( 1-F(r_{n}(x(y_u),x(y_v))) \big)\\
&=\prod_{u\in V}\nu(x(y_u))\prod_{(u,v)\in E}\sfrac
{F(r_{n}(x(y_u),x(y_v)))}{1-F(r_{n}(x(y_u),x(y_v)))}\prod_{(u,v)\in \skrie}
\big(1-F(r_{n}(x(y_u),x(y_v))) \big).
\end{align*}
Therefore, we have in the case of Theorem~\ref{randomgsmb.sparse}
\begin{align*}
 -\sfrac{1}{ n(\log(n))^2 }\log P(x)&=-\int_{\skrix}\sfrac{\log\nu(a)}{(\log n)^2}\, \skril_X^1(da) -\frac 12\,
\int_{\skrix^2}\,\sfrac{\log(F(r_{n}(a,b))/(1-F(r_{n}(a,\,b))))}{\log n}\,\skril_X^2(da,db)\\
&-\frac 12\,\int_{\skrix^2} \,\sfrac{\log(1-(F(r_{n}(a,b)))}{ (\log)^2/n
} \skril_X^1\otimes\skril_X^1(da,db)-\frac 12\, \int_{\skrix^2}
\,\sfrac{\log(1-(F(r_{n}(a,b)))}{(\log n)^2}\skril_{\Delta}^1(da,da).
\end{align*}

Now in the first case the integrands $\sfrac{-\log\nu(a)}{(\log n)^2},\, \,\sfrac{-\log(1-(F(r_{n}(a,b)))}{ (\log n)/n}\,\mbox{and}\,
\sfrac{-\log(1-(F(r_{n}(a,b)))}{(\log n)^2}$ all converge  to zero,
while $\sfrac{-\log((F(r_{n}(a,b)))/(1-(F(r_{n}(a,b)))}{\log n}
 \to 1,$   for  all  $a,b\in\skrix.$ Hence Theorem~\ref{randomgsmb.sparse} follows from
Theorem~\ref{WLLN}.



\end{document}